\documentclass[
    ,final            
  ,numberedheadings 
  ]
  {aipproc}

\newcommand{\bi}{\begin{itemize}}
\newcommand{\ei}{\end{itemize}}
\newcommand{\be}{\begin{equation}}
\newcommand{\ee}{\end{equation}}

\newcommand{\ben}{\begin{eqnarray}}
\newcommand{\een}{\end{eqnarray}}
\newcommand{\benstar}{\begin{eqnarray*}}
\newcommand{\eenstar}{\end{eqnarray*}}

\layoutstyle{6x9}

\begin{document}

\title{Single-domain protein folding: a multi-faceted problem}


\author{Ivan Junier}{
  address={Departament de Fisica Fonamental, Facultat de Fisica, Universitat de Barcelona, Diagonal 647, 08028 Barcelona, Spain}
}

\author{Felix Ritort}{
  address={Departament de Fisica Fonamental, Facultat de Fisica, Universitat de Barcelona, Diagonal 647, 08028 Barcelona, Spain}
}

\begin{abstract}
 We review theoretical approaches, experiments and numerical
simulations that have been recently proposed to investigate the
folding problem in single-domain proteins. From a theoretical point of
view, we emphasize the energy landscape approach. As far as
experiments are concerned, we focus on the recent development of
single-molecule techniques. In particular, we compare the results
obtained with two main techniques: single protein force measurements
with optical tweezers and single-molecule fluorescence in studies on
the same protein (RNase H). This allows us to point out some
controversial issues such as the nature of the denatured and
intermediate states and possible folding pathways. After reviewing the
various numerical simulation techniques, we show that on-lattice protein-like
models can help to understand many controversial issues.
\end{abstract}

\maketitle

\section{Introduction}
Electrostatic forces, Van der Waals interactions, hydrogen bonds
and entropic forces are the main elementary interactions that govern thermodynamics and
kinetics of molecular interactions. In solution, electrostatic forces
are mainly screened except at very short distances, typically on the
order of the Angstr$\ddot {\mbox o}$m. At these distances,
chemical bonds of
energies a few hundreds times larger than the thermal energy $k_B T$
tend to form at physiological temperatures --$k_B$ is the Boltzmann
constant and $T$ the temperature of the solvent. At physiological
temperatures, the free energy of the hydrogen bonds involved in the
formation of the protein secondary structures ($\alpha$-helix,
$\beta$-sheet) is around $2 \times k_B T$ \cite{finkelstein}. Also,
the Van der Waals potentials that are responsible for the protein
tertiary interactions are on the order of $k_BT$.  Thus, in proteins
(but also in nucleic acids), the thermal energy is comparable to the
free energy of formation of non-covalent interactions. This leads to
opposite effects.  On one hand, it means that thermal agitation is the
main source of intrinsic noise for biological processes
\cite{mcadams}.  On the other hand, it suggests that thermal energy
may be used as an energy source to trigger conformational changes and
therefore induce mechanical work at the molecular level
\cite{ritort}. However, in order to carry out specific tasks in a
highly fluctuating environment, evolution, through natural selection,
has favoured the formation of compact biological structures (DNA, RNA, protein) that are
stabilized by multiple non-covalent bonds.  RNAs and proteins are small
enough to be activated by a small amount of energy available from ATP
hydrolysis and, at the same time, stable enough to be biologically functional. 
DNA is a very long charged polymer but only a few number
of base pairs are involved during transcription or replication
processes.  Furthermore, proteins, such as DNA polymerases or
helicases, act {\it locally} on the DNA.

Proteins are ubiquitous molecules with a large variety of functions
(regulatory, enzymatic, structural,...) \cite{petsko}.  Regulatory
proteins are involved in gene regulation processes, structural
proteins (microtubules, actin filaments,....) give mechanical rigidity
to the cell, transmembrane proteins regulate ion and water transport
through membranes, etc...  Proteins do not usually work alone. In some
cases, a multiplex of several individual proteins participate in a
common task, such as helicases and DNA polymerase proteins that coordinate their action during
replication.  In other cases, proteins are subunits of large molecular
complexes such as the ribosome that consists of a patchwork of RNA and
protein subunits.

During cell activity, proteins
are continuously synthesized --and destroyed by protease proteins. Constitutive amino acids are transported
by the transfer RNA and the ribosomes synthesize polypeptide sequences by matching the genetic code
of the messenger RNA. Proteins have the remarkable ability to
{\it fold upon a native structure}. This propensity was demonstrated in 
{\it in vitro} experiments by Anfinsen {\it et. al} \cite{anfinsen}
 in a denaturation/renaturation experiment of the Ribonuclease A protein in presence of urea. 
Subsequently, it has become clear that this is a general property of proteins since many experiments on 
different proteins have led to the same conclusion \cite{dobson}. 
The fast folding property is crucial since a protein becomes active only by adopting a specific
thermodynamically stable structure (and in many cases by further forming specific complexes with small ligands). 
Furthermore, the diversity in protein functions
is related to the diversity of protein structures. 
Folding of large proteins 
is helped by specialized biological machines, the so-called chaperons (GroEL-GroES, DnaK-DnaJ,...).

The fast folding property is not trivial. A random sequence of amino acids (and in extreme cases a single amino acid mutation of
a good folder) leads to a polypeptide chain that behaves as a random coil without any
specific structure \cite{davidson,creighton}. To understand how proteins fold, 
different theoretical pictures have been proposed during the past twenty years \cite{finkelstein,dill,onuchic,thirumalai}. 
Interestingly, recent single-molecule experiments \cite{bustamante,bensimon,basche} 
that allow to investigate the biochemical processes at the molecular level \cite{bustamante}, in conjunction with
increasingly powerful simulations, have refuted some of the theories and sharpened the big picture. 
Such a symbiosis between experiments, theory and numerical simulations have led to a better understanding about
how biological machines work \cite{bustamante}. For instance, one can now think of models that predict 
the native 
structure of a protein from its primary sequence \cite{bradley}. At a different level, 
biophysicists are able to observe {\it in real time} the action of single proteins acting on DNA, 
such as Gyrase, a protein that relaxes DNA supercoils \cite{gore}, and study it under different conditions 
(temperature, pH, tension, torsion,...).

This article is a short review about the folding/unfolding of small proteins. We first discuss the theoretical ideas that are nowadays
used to tackle this problem. Next, we deal with the most recent experimental techniques that have provided important information about
the folding mechanism of different proteins. We finish by reviewing the
numerical techniques that are commonly used to investigate structural properties of the proteins during the folding transition. Our main goal 
is to present the basic notions necessary to 
understand the physics of the folding problem. The reader interested in a deeper understanding will find more detailed discussion
in the proposed references of each section.

In section \ref{FEP} , we discuss
the energy landscape picture, a useful scheme that provides an intuitive idea of the folding propensity \cite{dill,onuchic2} but also
that has led to quantitative tools useful to predict native and intermediate states \cite{onuchic}. We further
illustrate this approach
by describing two simple models that show a protein-like behaviour. In section \ref{SME},
we describe the main single-molecule techniques used to investigate individual proteins. We focus our
discussion in underlining the differences between single-molecule force and single-molecule fluorescence experiments. 
To this end, we compare studies that have been carried out with the protein RNase H. In section \ref{NS}, we review different
numerical techniques such as molecular dynamics. We explore the use of coarse-grained models and give more details
about generic lattice models that share protein-like properties. These models have the advantage of not being
time-consuming, and allow to tackle general properties expected in
single-domain protein folding. In particular, we address the questions of force-induced dynamics of a single-domain
protein.

\section{The free energy landscape picture}
\label{FEP}

\subsection{The Levinthal paradox}

\begin{figure}
\includegraphics[scale=0.3]{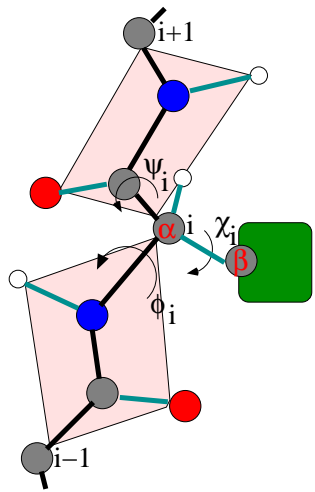}
\caption{A protein is a chain of amino acids linked by peptide bonds (the peptide units are outlined by he parallelograms).
The side-chain (outlined in green) defines the residue (amino acid $+$ side-chain). 
Twenty types of side-chain exist, the most simple being
an atom of hydrogen that is called glycine. In this case, there is no $\beta$-carbon. The peptide units are planar, 
due to the ${\mbox sp}^2$ hybridization type of the N-C bond. Different $\chi$ angles correspond to different conformations
of the side-chain. Two conformations ({\it cis} and {\it trans}) are possible 
for the peptide unit, depending on the positions of the oxygen and the hydrogen at the tops of the parallelogram.
Carbon, oxygen, nitrogen and hydrogen atoms are respectively represented in grey, red, blue and
white. The backbone structure is highlighted in black. 
}
\label{aminoacid}
\end{figure}

The structure of the native state of a protein is hierarchical. In the lowest level of description, a protein is described as
a sequence of amino acids (residues) linked by peptide bonds. There are twenty different types of
amino acids corresponding to different side chain groups --see Fig. \ref{aminoacid}. The residue sequence is called the 
primary structure. The formation of nearby hydrogen bonds between the amides and the carboxyl groups (Fig. \ref{aminoacid}) 
stabilizes the secondary structures mainly consisting of
$\alpha$-helices and $\beta$-sheets. The secondary structures are further stabilized by the
tertiary interactions that are either hydrophobic interactions or disulfide bonds.
Hydrophobicity results from the exposure of hydrophilic
side chains to the solvent leading to the condensation of polar residues inside the core of the protein.

\begin{figure}
\includegraphics[scale=0.2]{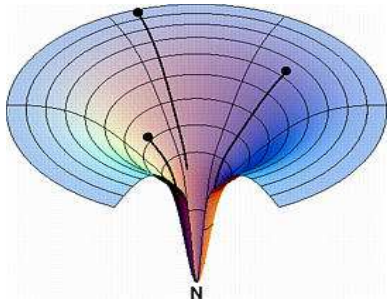}
\caption{Artistic cartoon of a perfect funnelled landscape. The vertical axis counts for the energy and the
horizontal plane for the degrees of freedom of the polypeptide chain. Taken from \cite{dill}.}
\label{funnel}
\end{figure}

Each individual peptide group can have two conformations (Fig. \ref{aminoacid}).
For an $M$-residue chain, one then roughly expects 
$2^M$ possible side chain configurations.
Assuming that the minimal timescale for a stereoisomeric conformational change is about
one picosecond, then the total time required for visiting all the configurations should be  
$ \sim 2^M {\rm ps} \sim 10^{10} {\rm years}$ for a $100$-residues protein \cite{finkelstein}. 
This crude approximation shows that the folding process can not consist of a random
search in the protein configurational space \cite{levinthal}. On the contrary, the {\it energy landscape}, 
i.e. the energy surface as a function of the configurational parameters (the degrees of freedom) 
--see Fig. \ref{funnel}-- ,
is biased toward  the native structure as depicted in Fig. \ref{funnel}.
Within the ideal picture of Fig. \ref{funnel},  
at sufficiently "low temperatures" 
(when $k_BT$ is on the order of the formation energy of a native contact \cite{zwanzig0}) the {\it free energy landscape}
 is biased by the energy gradient leading to downhill motion and collapse
towards the native structure \cite{ dill,onuchic,onuchic2}. This situation corresponds to a perfect
funnelled landscape \cite{onuchic}. The underlying mechanism that leads to such a smooth landscape is referred as
"minimal frustration" \cite{onuchic2}. A minimally frustrated structure is a structure for which the intra-molecular 
interactions are not in conflict
with each other leading to a smooth landscape as in Fig \ref{funnel}.  The 
concept of a funnel is not only qualitative but also quantitative. The simplest way to
design a perfect funnel is by considering interactions that only stabilize the native structure \cite{onuchic}. 
By following this strategy and
using a coarse-grained description of proteins (e.g. the G$\bar {\mbox o}$ model (see section \ref{go})), excellent
predictions of native structures and even intermediate states have been observed \cite{onuchic}.
However,  the perfect funnelled landscape is not so general.
A rough energy
landscape with many local minima and saddles corresponding
to configurations with various degrees of stability, is more appropriate. 
Within this picture, misfolded behaviour results from the competition between local minima that are close to the native state 
\cite{onuchic2}. For a detailed discussion about the energy landscape, see the review by Onuchic {\it et. al} \cite{onuchic2}.

\subsubsection{Mixing stochasticity and determinism}

The energy landscape picture, by definition, leads to non-specific folding pathways from the many denatured 
(i.e. non native) conformations 
to the folded native state. This scheme has been opposed for a long time to the very first scenarios aiming at
explaining the folding property.
According to the latter, the folding process is
a specific mechanism whose dynamics is sequential, which leads to a unique folding pathway \cite{finkelstein}, 
by opposition to the stochastic nature
of the energy landscape approach \cite{dill,onuchic2,finkelstein}. These two view points are not contradictory but rather describe
mechanisms at different levels. For instance, Lazaridis and
Karplus \cite{lazaridis} have studied 24 unfolding trajectories 
of a small protein (chymotrypsin inhibitor 2), using molecular dynamics. They have observed
large statistical fluctuations in the gyration radius of the successive structures 
during the unfolding process, in agreement with the energy landscape picture. On the other hand,
 some specific events, such as the destruction of
tertiary contacts, were found to be specifically ordered in time \cite{lazaridis}.

\subsection{Thermodynamics}

In most cases, small globular proteins fold following an all-or-none process, just as do small RNA hairpins.  
The origin of this cooperative effect lies in the fact that the native state has a very low entropy.
Thus, the transition from the denatured state (with high entropy) to the native state is generally accompanied by
an entropy jump or, equivalently, a peak in the specific heat as observed in bulk denaturation experiments \cite{finkelstein}.
In the following, we discuss two simple models describing the transition between a high entropy phase and a very low
entropy native phase.

\begin{figure}
\includegraphics[scale=0.25]{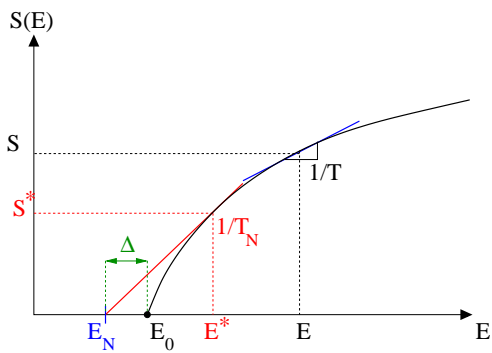}
\caption{Dependence of the entropy in a random energy protein. In the pure random energy model (without the
native state at energy $E_N$), the mean value of the
energy $E$ and the entropy $S$ are given by the coordinates at which  
the derivative of $S(E)$ is equal to $1/T$. In the composite model (i.e. with the native state), 
the free energy equality between the native state and
the denatured states implies $1/T_N=S^*/(E^*-E_N)$. $T_N$ is the transition temperature where the native state
and the denatured state are equal likely. $\Delta$ is the energy gap.}
\label{REM}
\end{figure}

\subsubsection{The entropy crisis avoided.} In the glass phenomenology, it has been hypothesized that there is a finite
temperature at which the configurational entropy ([total entropy]$\; -
\;$[vibrational entropy]) vanishes \cite{ediger}.  This has been
called "the entropy crisis" by Kauzmann \cite{kauzmann}. The
simplest model describing the entropy crisis is the random energy
model (REM) \cite{derrida}. In this model, the entropy is a quadratic
function of the energy that vanishes at an energy $E_0$, i.e.
there is no state with an energy below $E_0$. At equilibrium, the
free energy corresponds to the point in the entropy curve, $S(E)$, at
which its tangent is equal to $1/T$ --see Fig. \ref{REM}. As a
consequence, as $T \to T_0$ where $T_0$ corresponds to the energy
$E_0$, the entropy {\it continuously} vanishes. The point at $E=E_0$
defines the glass transition. By incorporating into this model a
native state with an energy $E_N=E_0-\Delta$ ($\Delta$ is the
so-called energy gap), one gets a first-order transition
\cite{onuchic2}, between a high entropy state and the native state, at
a temperature $T_N=(E^*-E_N)/S^*$. At $T_N$, the free energy of the
denatured state, $F^*=E^*-T_NS^*$, is equal to the native one,
$F_N=E_N$. It can be shown that the glass transition is avoided if the
energy gap is much larger than $E_0/M$
\cite{bryngelson,onuchic,finkelstein}, $M$ being the number of
residues.  The transition then becomes first order.

\begin{figure}
\includegraphics[scale=0.2]{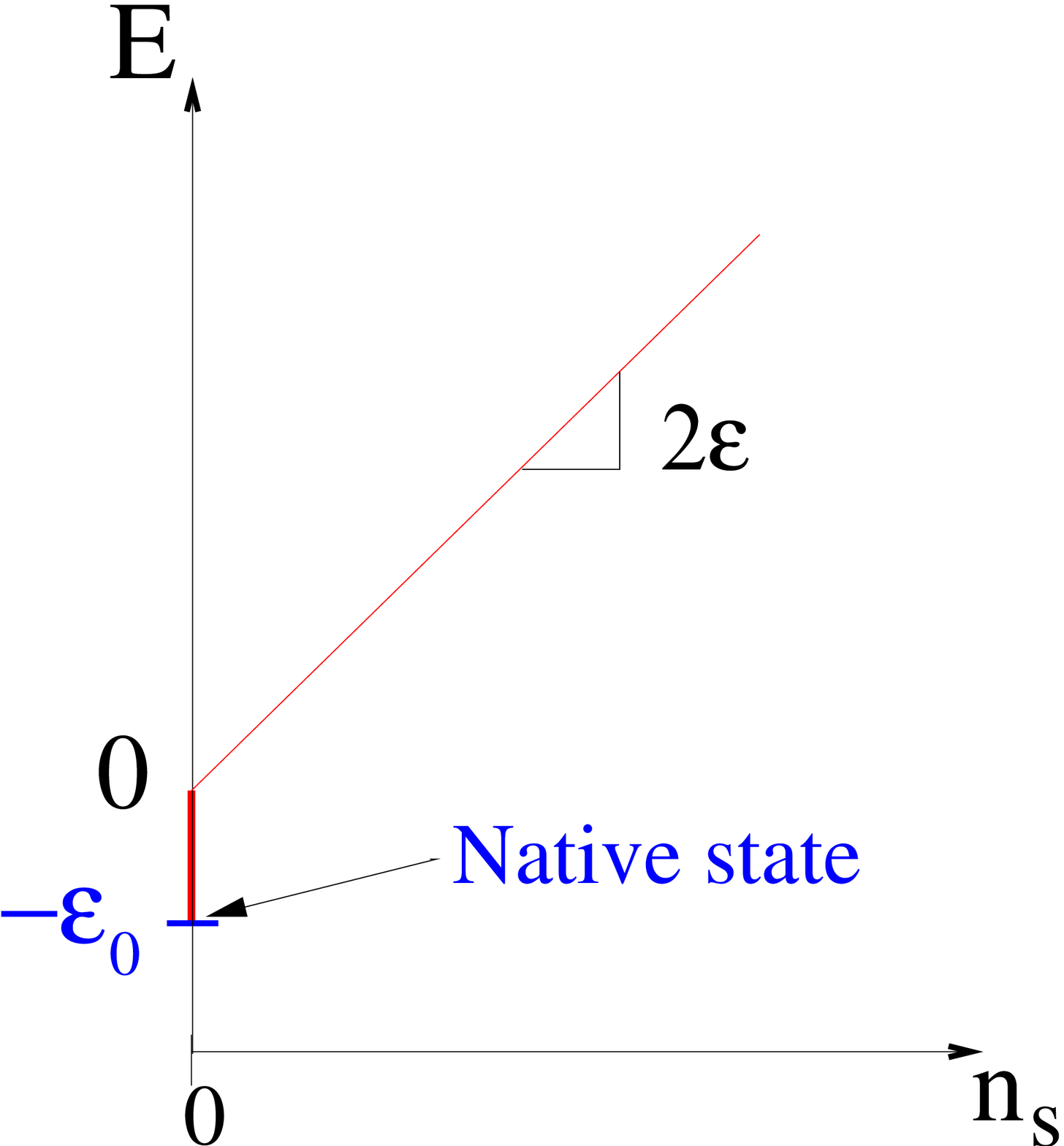}
\includegraphics[scale=0.2]{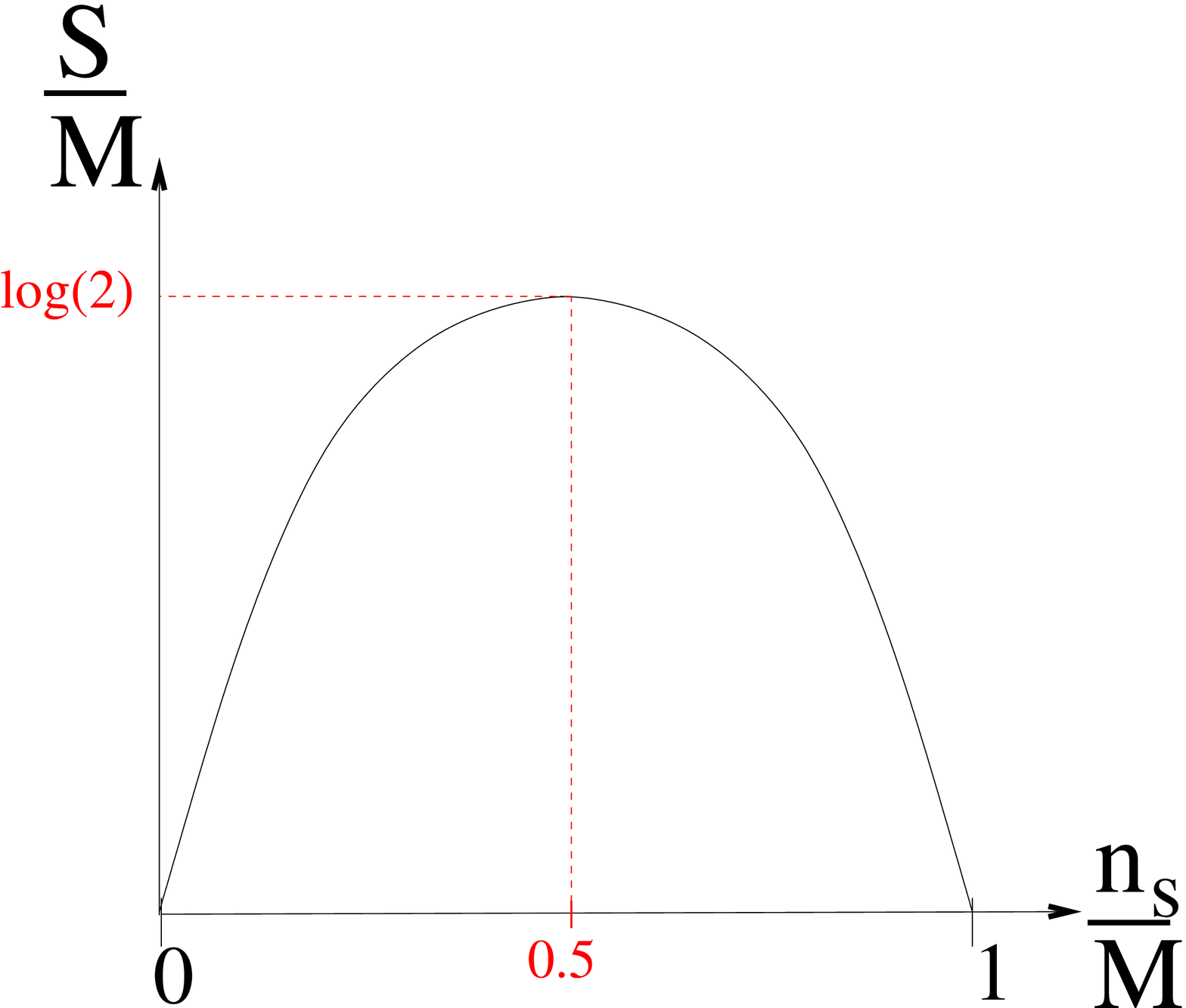}
\caption{The Zwanzig picture. $n_s$ is the number of non-native contacts. Left: the potential $E(n_s)$ is given by $E(n_s)=2\epsilon n_s$
when there are non-native contacts ($n_s>0$). 
At the native state, $E(0)=-\epsilon_0$. Therefore, the energy gap between the native state and
the lowest denatured states is equal to $\epsilon_0$. Right: the entropy as a function of $n_s/M$ for large $M$. In this limit,
$S(x=n_s/M)/M=(x-1)\log(1-x)-x\log x$.}
\label{zwanzig}
\end{figure}

\subsubsection{Funnel-driven transition}
The energy funnel is akin to the minimal frustration property.
The simplest model exhibiting a perfect funnelled landscape is the Zwanzig model \cite{zwanzig}. 
It can be thought of as a spin model where the energy of a given spin configuration $\mathcal{C}=\{s_1...s_M\}$ (we
consider a set of
$M$ spins) reads 
$E=\epsilon\sum_i |s_i-s^{N}_i|/2-\epsilon_0\delta(\mathcal{C}-\mathcal{C}^{N})$ where 
$\mathcal{C}^{N}=\{s^{N}_1...s^{N}_M\}$ is the native configuration. The parameters $\epsilon$ and $\epsilon_0$
are positive energies related to the gradient of the funnel and the native gap $\Delta$ respectively --see Fig. \ref{zwanzig}.
The energy of this system can be explicitly written as a function of 
the number $n_s$ of spins that differ from the spins 
in the native configuration \cite{zwanzig}. We will call $M-n_s$ the number of native contacts. 
Let us now consider a single-spin dynamics with Metropolis rules. In this case, at any time there are only two kinds of
elementary moves: a spin-flip can lead to a new native contact ($n_s \to
n_s-1$) or to a new non-native contact ($n_s \to n_s+1$). Since there is no interaction between the spins, there is
no conflict between the interactions, which means that the probability to have a native contact at a site $i$ does not 
depend on the configuration of
the other spins. A set of non-interacting constituents 
that
feel a time-independent local potential is therefore the simplest example of a minimally frustrated system since there is no
frustration at all.

One can 
write a master equation for the probability density of the number of contacts $n_s$ at a given time, 
where $n_s$ represents a 
reaction coordinate \cite{zwanzig}.
The thermodynamic potential $E(n_s)$, that reflects the funnel-shape, is linear in $n_s$ and has a gap at $n_s$=0 
(Fig. \ref{zwanzig}).
As occurs in any folding transition, there is a competition between the entropy, that favours 
denatured states (non-native contacts), and the potential energy $E(n_s)$ that biases the system towards the native structure. 
In the limit of large number of spins, one finds a
temperature transition, $T_N=\epsilon$, at which the probability of being in the folded/unfolded state is $1/2$ \cite{zwanzig}.

\begin{figure}
\includegraphics[scale=0.35]{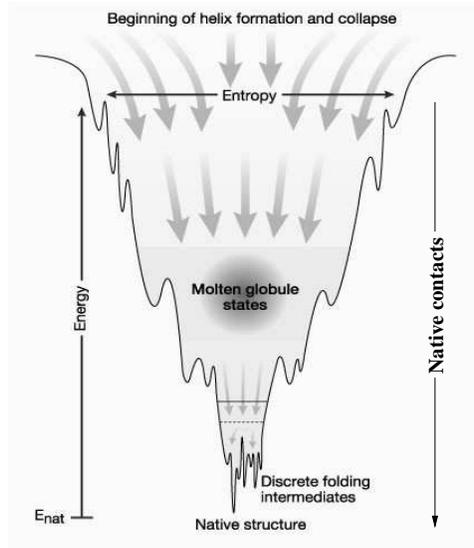}
\caption{The energy landscape picture. In general, the energy
landscape of a protein is rugged and funnelled, i.e. with an overall
gradient that is oriented toward the unique native state. The
landscape is stratified according to the energy of the configuration,
or according to the percentage of native contacts. Misfolded states
are the local minima closed to the native state. Adapted from
\cite{onuchic2}.}
\label{wolynes}
\end{figure}

\subsubsection{The potential of mean force}

Taking into account the ruggedness of the energy surface, the
funnelled shape of the {\it potential energy landscape} is usually
represented as in Fig. \ref{wolynes} \cite{onuchic2}. However, the
{\it free energy landscape} is more suitable to discuss a possible
thermodynamic transition.  Contrary to bulk experiments, in which
measurements lead to ensemble averaged quantities, single-molecule
experiments allow to compute the free energy as a function of reaction
coordinates such as the molecular extension, a quantity that is
related to the number of native contacts. This is usually called the potential
of mean force.  In the following, we discuss situations where the free
energy is projected along a single reaction coordinate. Nevertheless,
it must be stressed that many aspects of the folding kinetics can not
be understood without considering more than one reaction coordinate
(see e.g. \cite{leeson}).

\begin{figure}
\includegraphics[scale=0.3]{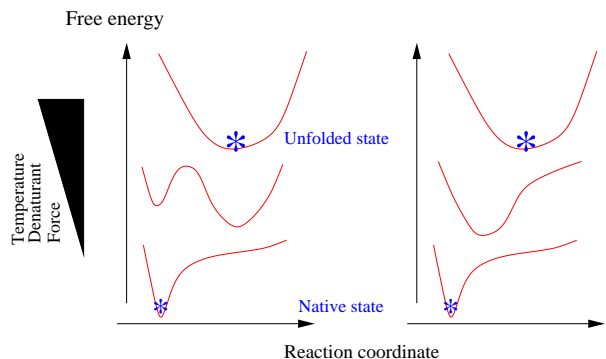}
\caption{The free energy landscape, or equivalently the potential of
mean force. Two scenarios are usually observed. Left: the transition
is first order with a coexistence phase between the native state and a
denatured state. Right: The transition is continuous. The free energy
projection always shows a single well that drifts toward the native
state as the folding conditions become more appropriate (bottom
panels). The $*$ indicate the denatured states and the native states.}
\label{free}
\end{figure}

\begin{figure}
\includegraphics[scale=0.25]{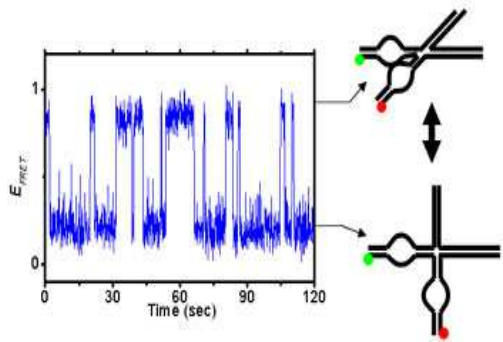}
\caption{Example of cooperative transitions observed in a single-molecule experiment \cite{tan}. Left: the vertical axis represents the
FRET efficiency (see section \ref{secFRET}) that reflects the state of the biomolecule (here an RNA molecule hairpin ribozyme shcematically
represented on the right part of the figure). One can see that
the molecule switches between the native state (upper configuration on the right)
and the denatured state (lower configuration). Taken from \cite{ha}.}
\label{twostate}
\end{figure}

The free energy landscape approach calls for two general situations (Fig. \ref{free}) that have been experimentally observed. 
On one hand, some single-domain protein experiments have revealed
a continuous phase transition between denatured states and the native state \cite{munoz}. 
In this case, at any temperature, the free energy landscape
is composed of a single well. The minimum of the well drifts towards the native structure as one lowers the temperature 
or decreases the denaturant concentration. On the other hand, most of the single-molecule (and bulk) 
experiments involving single-domain proteins 
have revealed a first order transition \cite{finkelstein,baker}: 
the free energy profile consists of two wells with minima corresponding to the denatured and the
native states --see Fig. \ref{free}. 
As in any first-order transition, the system goes from a denatured state to the native state by passing through a coexistence
phase. In bulk experiments, this leads to the presence of proteins that are denatured and proteins that are in a native state. 
From the single-protein
point of view, this suggests cooperative switches between the native and the denatured states as reported in Fig. \ref{twostate}.

\subsection{Kinetics}

The energy landscape represents a useful picture to understand the existence of misfolded structures and, 
more generally, the
folding kinetics of a single-molecule. By considering the overdamped motion
of a particle along a potential of mean force (free energy projection), one implicitly makes the assumption that all 
the degrees of freedom
orthogonal to the reaction coordinate locally equilibrate \cite{risken}. This may not be true in specific non-equilibrium
conditions, such as low temperatures where proteins show glassy behaviour \cite{onuchic2}.

\subsubsection{Two-states and downhill kinetic scenarios}
\label{2s}

Kramers theory allows to derive dynamical properties related to the diffusion motion of a particle
along a one-dimensional landscape \cite{zwanzigbook}. In particular, 
the mean first-passage time between the denatured and the native states can be computed to extract
the effective free energy barriers. 
A two-states description of the Kramers' problem models the dynamics in terms of activated events across 
a free energy barrier and represents the
simplest description of a cooperative all-or-none transition. This approach is potentially useful 
to understand single-molecule force experiments, e.g. in the force unfolding of single RNA molecules 
\cite{ritort2}. When the position of the transition state 
moves along the reaction coordinate by changing the external conditions 
(temperature, denaturant concentration, stretching force),
known as the Hammond behaviour \cite{hammond},
an extended two-states description with a mobile
barrier can be applied \cite{manosas}. 

The existence of free energy barriers
that make the transition all-or-none is usually attributed to the asynchronous compensation between 
energy gain and entropy loss \cite{onuchic2}.
However, continuous transitions have been also observed in recent experiments \cite{garcia}.
These transitions can then be thought as a limiting case of the two-states model where the free energy 
barrier becomes comparable to $k_BT$. It is then more convenient to see the folding as a downhill process \cite{munoz,garcia}.
Notice that the ideal funnel picture of Fig. \ref{funnel} 
actually suggests a compensation between entropy (given by the radius of the funnel) and energy (given by the
depth of the funnel).

The two-states transition between either the native state and a random
coil (with no native contacts) or the native state and a
molten-globule structure (with numerous native contacts) has been for
long time a well accepted scenario for single-domain proteins
\cite{finkelstein}. However, this has been disputed in recent
numerical studies on the lyzozyme (1HEL) \cite{fitzkee}. This is a
single-domain protein known to exhibit at room temperature a first
order transition between the native state and a pure random coil as
the concentration of denaturant (guanidine dihydrochloryde) is
increased \cite{tanford}. In fact, it has been shown that even in the
presence of many native contacts (more than $90\%$) the gyration
radius and the end-to-end distance are well described by the Gaussian
random coil model \cite{fitzkee}. Therefore, standard bulk experiments
may not provide enough information to distinguish a
random coil state from a native-like state. In the next section, we
describe single-molecule techniques that might resolve this
controversy by addressing new interesting questions.

\begin{figure}
\includegraphics[scale=0.2]{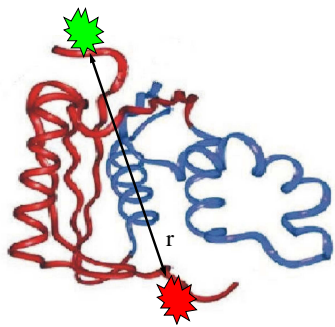}
\caption{The RNase H protein structure. The colourful stars represent the dyes that are chemically attached to the protein. 
These dyes are used in the florescence techniques, namely the FRET measurement --see text. $r$, the distance between the dyes, 
is directly related to the molecular extension of the protein. Adapted from \cite{cecconi}.}
\label{RNaseH}
\end{figure}

\section{Single-molecule experiments}
\label{SME}

In this section, we review the principal techniques used to investigate proteins 
 in single-molecule experiments. We focus our discussion on two of  
them: fluorescence spectroscopy and force measurements. We compare the results obtained
with these techniques in the RNase H protein and discuss whether
force may induce folding pathways different than those of thermal folding.

\subsection{Fluorescence techniques}
\label{secFRET}

Three-dimensional native structures can be determined in solution 
by nuclear magnetic resonance (NMR) spectroscopy or in crystal forming proteins by X-ray crystallography. For small globular
proteins, the two measurements give generally the same result, showing that the native structure is a highly
compact structure in solution. Such techniques are inappropriate for studying the structure of the transition state and 
the denatured states. Indeed, it is impossible to crystallize fluctuating states and
NMR measurements average out conformational fluctuations. 
Nevertheless, some bulk techniques, e.g. small-angle
X-ray and neutron scattering, have provided precious information about
 quantities such as the radius of gyration \cite{millett}. In particular, it has been shown that random coils are not the most
 general denatured state \cite{millett}, even at high denaturant concentrations. Recently, these results have been 
unambiguously confirmed
 by using single-molecule fluorescence techniques.

Fluorescence techniques are based on the so-called F$\ddot {\mbox
o}$rster resonant energy transfer (FRET).  A green fluorescent donor
dye and a red fluorescent acceptor are chemically attached to the end
residues of the protein (Fig. \ref{RNaseH}).  The donor is excited by
a well-tuned laser and further relaxes by emitting a fluorescent light
that can be detected by a spectrophotometer. The acceptor is chosen
such that its absorption spectrum overlaps the emission spectrum of
the donor. As a consequence, a non-radiative energy transfer between
the chromophores may decrease the intensity of the donor by enhancing
the emission of the acceptor. The (FRET) efficiency of energy transfer
between acceptor and donor depends on their distance $r$, and hence on
the protein extension, through the simple relation 
\be
E=\frac{1}{1+(r/R_0)^6}
\label{ef}
\ee
where $R_0$ is a characteristic parameter of the pair of dyes.
On the other hand, the efficiency $E$ can be directly 
related to the emitted intensities by the dyes:
\be
E=\frac{I_A}{I_D+I_A}
\ee
where $I_D$ and $I_A$ are the intensities emitted by the donor and the acceptor. 

As a result, a quantitative spectral detection
of the dyes gives information about the conformation of the protein. One may even think of placing
the dyes at different locations in the protein to get further structural information. In this spirit, this technique has been
recently used to investigate specific conformational changes during biological processes. For instance, 
it has been used to follow the different steps of protein synthesis in the ribosome \cite{blanchard}. This is very important
to understand the mechanism responsible for the so exclusive codon/anticodon recognition by the transfer RNAs.

\begin{figure}
\includegraphics[scale=0.3]{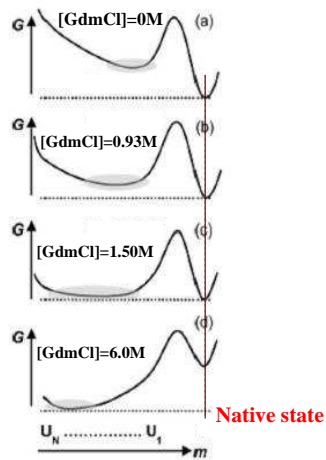}
\caption{Free energy profiles of the RNase H as a function of the
denaturant concentration $[GdmCl]$. Upper panel: the thermodynamically
stable state is the native state. Bottom panel: the thermodynamically
stable state is a denatured state. The abscissa is the so-called
cooperativity parameter and is related to the propensity of the state
to let the solvent enter into the molecule. $\{U_i\}_{i=1…N}$ is a
set of denatured states structurally close to the native state, the
closest being $U_1$. Taken from \cite{nienhaus2}.}
\label{nienhaus}
\end{figure}

\subsubsection{The RNase H protein}
Let us now focus on folding studies in a small single-domain protein, the 155-residue RNase H protein.
The native structure of this protein is well known and is shown in Fig. \ref{RNaseH}. Under appropriate folding
conditions, several (bulk) studies have shown that the folding is preceded by a fast collapse to a compact structure presumably 
stabilized by a central nucleus \cite{raschke,baldwin}.

Nienhaus and co-workers have used the above fluorescence technique to investigate several structural properties of the RNase 
H protein 
\cite{nienhaus,nienhaus2}.
To determine the
spectral properties of the dyes, they fix an ensemble of proteins on a glass surface. 
A FRET histogram is obtained 
by "counting" the number of proteins with efficiency $E$ (see Eq. \ref{ef}). By varying the concentration of denaturant, 
e.g. the guanidine dihydrochloride, they have monitored the cooperative transition between the native
and the denatured states. From these measurements, it is then possible to extract 
the corresponding potential of mean force (free energy landscape) along a reaction coordinate
that is related to the compactness of the protein --see Fig. \ref{nienhaus}. This coordinate actually
characterizes the propensity
of the molecule to let the solvent enter.
Notice that these curves could be the curves of any single-domain protein. Interestingly, 
the folding free energy changes as one varies the concentration of denaturant. This raises several questions: 
to what extent the denatured state at low denaturant concentration is different from the denatured state at high concentration? 
Is the transition
between the high-denaturant state and the low-denaturant state of the same type 
as the continuous transition discussed above? Does the high concentration denaturant state have a residual
structure reminiscent of the native state?
Fluorescent studies by the Nienhaus group have shown that even at high denaturant concentration, the denatured 
state was composed of non-random structures \cite{nienhaus,nienhaus2}. This study, in conjunction with the numerical simulations of
the folding of lyzozyme (1HEL) \cite{fitzkee}, cast serious doubts about the true nature of the denatured state, an issue
 that has been an experimental challenge for a long time. 
Moreover, it is also a difficult problem from the point of view of numerical simulations because of the 
huge number of accessible configurations.

The RNase H results \cite{nienhaus} suggest the existence  at low denaturant 
concentration of a well-defined compact structure 
different from the native state. As discussed above, this structure was expected from earlier stop-flow 
kinetic experiments
in which RNase H often showed the accumulation of compact structures during the dead-time of the measurement. 
Interestingly, force experiments applied to the same molecule have also shown the existence of a well-defined 
intermediate state that
coexists with both the denatured and the native states \cite{cecconi}.

\subsection{Force measurements}

Force measurements on a single molecule have been first realized on a
double-stranded DNA \cite{bustamante}. A fluid flow and a magnet were
used to stretch the molecule that was attached to micron-size
beads. The measurement of the molecular extension of DNA has then
revealed unexpected mechanical properties, such as the overstretching
transition \cite{bustamante}. Subsequently, different studies have
been realized \cite{bustamante,bensimon} in order to investigate the
behaviour of DNA under torsional strain (using magnetic beads that can
be rotated by magnets) \cite{strick,strick2}, the DNA and RNA
unzipping process (using optical tweezers) \cite{bockelmann,cocco},
the DNA packaging problem \cite{smith}, DNA/protein interactions
\cite{allemand} or DNA condensation \cite{ritort3}. As a consequence, single-molecule force
experiments have contributed a lot toward our understanding of the
cell machinery.  Single-molecule force measurements (on RNA)
\cite{collin} have been also used to test non-equilibrium theories in
statistical physics and to recover folding free energies in RNA
molecules.  In this spirit, biomolecules appear to be ideal systems to
explore the thermodynamic behaviour of small systems and to test
non-equilibrium theories in statistical physics \cite{bustamante2}.

Recent nano-manipulation of single protein molecules using the atomic force microscope (AFM) \cite{fisher2} have provided
direct evidence for sequential unfolding of individual domains upon stretching \cite{fisher,kellermayer,rief,schlierf}. However,
optical tweezers are more appropriate to study the unfolding/folding dynamics of small single-proteins (and 
small RNAs). In fact, the folding free energies of such biomolecules
are on the order of $100 k_B T$ at room temperature.
Considering a typical gain in extension of $\Delta x \sim 10-20 nm$, between the native state and a stretched state, 
a mechanical 
energy $f \Delta x \sim 100 k_B T$ is provided by a stretching force on the order of $10 pN$, which is in the 
ideal working range of 
optical tweezers \cite{lang}. In contrast, the AFM technique is useful to investigate forces above a few tens
of $pN$ \cite{fisher2} but can not reach forces $\sim pN$ mainly because of the high spring constant of the cantilever.

\begin{figure}
\includegraphics[scale=0.3]{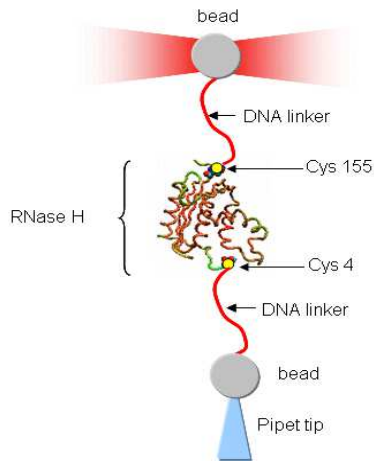}
\caption{
Setup of the force measurement in the single-protein RNase H experiment. DNA linkers between the beads and the protein 
are inserted in order to be able to manipulate the protein. The DNA linkers are chemically attached to the end of the 
proteins via the insertion of a cystein side-chain. The bead at the top of the pipette is held fixed by air suction and 
the other bead is trapped in the optical well.}
\label{setup}
\end{figure}

\begin{figure}
\includegraphics[scale=0.3]{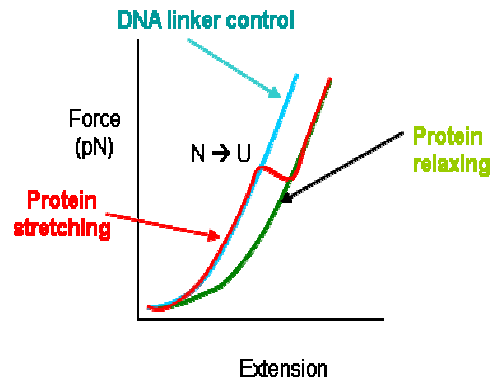}
\caption{Typical force-extension curve (FEC) during the unfolding/refolding ramp force protocol of a single-domain 
protein. The undolding of the protein corresponds to the extension jump (in red). The rest of the curve is well 
described by a worm-like chain model that models the extension of the linkers when the protein is still folded, and 
the extension of the (linkers $+$ unfolded protein) when the protein is unfolded.}
\label{FEC}
\end{figure}

Typical optical tweezers experiments use micron-sized glass chambers filled with water and two beads. The protein is
chemically labelled at its end and polystyrene beads are chemically coated to stick to the ends
of the labelled molecule. Because proteins are too small to be manipulated with micro-sized
beads, a tether consisting of a double stranded DNA
is inserted between the beads and the molecule that acts as a polymer spacer --see Fig. \ref{setup}. 
This prevents Van der Waals forces between the beads and the protein and
allows a direct manipulation of the protein. One bead
is then held fixed by air suction on the tip of a glass micro-pipette, the other is trapped in the focus of
a laser beam. When the bead deviates from the focus a restoring force acts upon the bead, the principle being the same
by which a dielectric substance inside a capacitor is drawn inwards by the action of the electric field. To a good approximation,
the trap potential is harmonic. Thus, knowing the trap stiffness, it is possible to apply mechanical
force (by moving the bead) and to observe in real-time the force-extension curves (FEC). In the FECs, the force
acting on the molecule is represented as a function of the end-to-end distance between the two beads. The cooperative
opening of the proteins is characterized by a jump in the extension of the molecule --see Fig. \ref{FEC}. By studying the stochastic
properties of the FECs, one is able to recover the distance from the native state to the transition state and map the
free energy landscape as a function of the molecular extension. The folding and the unfolding rates can
also be determined \cite{manosas,schlierf}.

\subsubsection{The RNase H protein}
\begin{figure}
\includegraphics[scale=0.25]{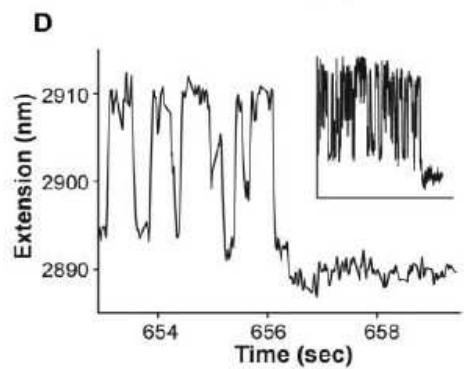}
\caption{Extension trace of the RNase H protein at constant force ($ \approx 5.5 pN$). In the first part, we see successive 
all-or-none transitions between the intermediate state and the unfolded state (fully stretched). Then, a transition 
occurs between the intermediate state and the native state showing that the intermediate state is on-pathway. Figure taken from \cite{cecconi}.
}
\label{interme}
\end{figure}

\begin{figure}
\includegraphics[scale=0.2]{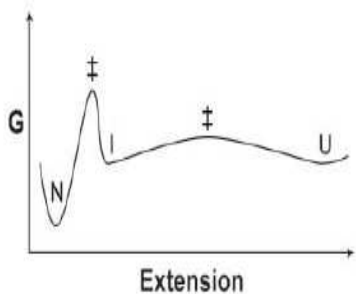}
\caption{Free energy (G) in RNase H protein projected along its molecular extension. Three regions, that are delimited 
by the signs $\ddagger$, can be defined: the native region where the minimum corresponds to the native state (N),
the intermediate region where the minimum corresponds to the intermediate state (I) and the stretched region
where the minimum corresponds to the unfolded state (U).
}
\label{freeinterme}
\end{figure}

In the case of the RNase H protein, Cecconi {\it et al} \cite{cecconi}
 have shown that mechanical forces can stabilize an intermediate
 state. We use the word "stabilize" since the intermediate state
 corresponds to a local minimum of the free energy landscape projected
 along the end-to-end distance (Fig. \ref{freeinterme}) that is well
 separated from the unfolded state and the native state by all-or-none
 transitions (Fig. \ref{interme}).  At constant force, three regimes
 can be distinguished depending on the value of the force: at high
 force, the molecule is fully stretched and no native residual
 contacts are present; at low force, the protein is in its native
 compact state; in-between, there is an intermediate state with a
 partial number of native contacts that are formed.  Three states, instead of two,
 coexist: the stretched, the native and the intermediate compact
 states (see Fig. \ref{freeinterme}). A statistical analysis of the
 breakage force and measurement of the rip extensions have led to an
 extrapolated zero-force intermediate free energy that correlates well
 with that of the early compact structure that forms in bulk
 experiment \cite{raschke}.

\subsection{Comparing force and FRET measurements}
A comparison between the folding/unfolding study of RNase H {\it with} and {\it without} force 
raises interesting questions relative to the structure of proteins: 1) Under which conditions do we expect
that
the early molten-globule state that forms at zero force is the intermediate state stabilized by
mechanical force? 2) FRET measurements have revealed a hierarchical structure of RNase H in the denatured state 
\cite{nienhaus2}. Is the stabilization of the intermediate state related to this observation? 3) More generally,
is the stabilization of an intermediate state a signature of a specific folding mechanism?
Such questions can be actually addressed in numerical simulations 
of simpler models \cite{junier}. 

The force measurements in RNase H also raise questions about the on/off pathway nature of the intermediate states, 
an issue that we discuss in the next paragraph.

\subsection{Probing the nature of the intermediate states}
\label{inter}

Let us consider a system with a free energy landscape showing three well-separated
minima --see Fig. \ref{freeinterme}. One might wonder whether a diffusive dynamics along this profile 
fairly reproduces the observed dynamical behaviour in the single-molecule experiment (Fig. \ref{interme}). In such case,
by starting from any state in the intermediate region and preventing the system from
going to the stretched region, the molecule should be able to fold to the native state. 
Misfolded (off-pathway) states are those that can not lead to the native state without unfolding back to the stretched states. 

In the RHase H force study, the extension trace of Fig. \ref{interme} suggests that folding indeed takes place {\it via} 
the intermediate
states \cite{cecconi}. However, we can not discard additional states lying at the same coordinate than the intermediate state 
and that can not
lead to the native state without unfolding back to the stretched state.
We can propose an
experimental protocol to quantify the fraction of off-pathway states with respect 
to on-pathway states.
Each time the system jumps to the intermediate state, we suddenly relax the force to a lower value. 
On-pathway states should quickly
lead to the native state without unfolding back to the stretched state. For off-pathway states however,
it is expected a first unfolding event to the stretched state and, most likely, an extremely large  folding time 
as compared to the typical on-pathway folding time. We give a numerical example of such a protocol in the next section.

\section{Numerical simulations}
\label{NS}

From an experimental point of view, it is still a challenge to get atomic structural information of
intermediate states, transition states  (corresponding to the maxima of the projected free energy landscape), or denatured
states. One
could think of a fluorescence technique using dyes attached to different residues of the proteins. 
However,
the presence of the chromophores inside the molecule is likely to impede the correct folding of the molecule or to
modify the real structure of the expanded states. So far,
the best way to characterize non-native states has been to resort to numerical simulations. The latter can be divided 
into three classes:
\begin{enumerate}
\item{\it Molecular dynamics.}  It takes into account all (or almost)
the atomic details of the molecule and the solvent can be explicitly
or implicitly treated \cite{karplus}. The folding pathways are
determined by simulating trajectories in quasi reversible conditions.
The technique is time limited because only nanosecond long unfolding
trajectories can be obtained whereas folding of real proteins occurs
mostly in microsecond timescales.  High temperatures or mechanical
forces are then usually used to accelerate the unfolding trajectories
\cite{karplus}. The main argument is that different conditions induce
different timescales but not different mechanisms.  Current
improvements in this field have been achieved thanks to the
development of more and more accurate interatomic potentials in
different environments \cite{cornell}.

Interestingly, the original technique has been also adapted to
simulate a large set of short trajectories (starting from random configurations) of a designed small protein (23 residues) 
at room temperature \cite{snow}.
A non-negligible amount of very fast folding trajectories ($\sim 20 ns$) has been observed whereas 
in experiments the mean folding time is on the order
of microseconds. This shows, as expected for two-state cooperative proteins, 
that the folding step
is very short but the whole folding mechanism is slowed down due to the presence of many possible denatured configurations. 
As mentioned in section \ref{2s}, this comes from an asynchronous compensation between entropy and energy.

\item{\it Coarse-grained models.}  Within this scheme, one reduces the
all-atom description to a mesoscopic description by neglecting
details of the polypeptide chain \cite{baker,zhou,veitshans}.  The
parameters of the mesoscopic description are obtained by a close
comparison with experiments.  Different levels of simplification are
usually taken into account. Perfect funnel models including only
interactions that stabilize the native structure have led to excellent
predictions of native and transition states \cite{baker,onuchic}.
Less restrictive G$\bar {\mbox o}$ models lead to a more refined
statistical description of folding trajectories
\cite{clementi,onuchic}.  At the end of this section, we describe
details of such simulations and discuss the issue in the presence of
mechanical forces.  Notice that modelling of the solvent is also
essential to understand protein folding.  Within this scope, G$\bar
{\mbox o}$-like potential including a (de)solvation potential can be
taken into account to model the expulsion of water molecules from the
protein core \cite{onuchic}.

\item{\it Protein-like models.}
The purpose of these models is not to study the folding mechanism of some specific proteins but rather to give general insights about
the folding dynamics. They are used to determine whether the dynamics is related to 
the (hetero-)polymeric properties of proteins, such as the native state geometry or 
the contour length of the chain.  They are also useful to investigate the folding behaviour in presence of specific 
external conditions (temperature, denaturant concentration, 
stretching
force...). Although there may have qualitative differences between on-lattice and off-lattice models (see e.g. \cite{pande}), 
most of the studies have been done with models defined on a lattice, the main reason being the possibility to simulate large molecules during long times.
In the following, we review some of these models and show how to incorporate mechanical forces. In particular, we show that
these models can be useful to tackle the problem about the on/off pathway states.
\end{enumerate}

\subsection{Protein-like behaviour of simple models}
\label{go}

\subsubsection{Hydrophobic-Polar models}

Heteropolymers on a lattice with simple hydrophobic-polar interactions
between non-adjacent monomers are the simplest models that show a
protein-like behaviour.  In the $HP$ model \cite{chan,chan2}, a
diblock copolymer chain composed of hydrophobic ($H$) and hydrophilic,
equivalently polar, ($P$) monomers is considered on a square lattice.
Only the interactions $HH$ are energetically favourable, the so-called
G$\bar {\mbox o}$ interaction.  Specific sequences (HPPH...) then lead
to protein-like behaviour and have been used to exhaustively explore
the underlying energy landscape \cite{chan2}. Interestingly, the
folding mechanism has been shown to be reminiscent of small
single-proteins. Indeed, under appropriate folding conditions, the
extended chain quickly condensate into a rich $HH$-bonds structure.

Since non-native $HH$ bonds are present,
the molecule further needs to break $HH$ bonds to get closer to the native state. This stage is similar to the exploration
of a non-native compact structure set that precedes the fast downhill step. Such a process actually goes accompanied 
with an expansion
of the structure in order to allow local conformational changes of the polymer, a behaviour that has been experimentally observed 
\cite{chan2}.
More quantitatively, a recent study of this model \cite{kachalo} has
pointed out that the folding rates may not be correlated to the thermodynamic properties of the molecule, such as the value of
the energy gap
and the structure of the native state. It rather suggests that the folding rates 
are well correlated with the number of local energy minima, i.e. the former decreases as the latter increases.
These results are in good agreements with some recent experiments \cite{scalley} but disagree
with other experiments that have shown a correlation between the
native structural properties and the folding rates \cite{plaxco}.

Notice that this kind of models do not present an intrinsic
hierarchical structure (primary, secondary and tertiary) as in
proteins. They should be rather thought of as a rough modelling of a
mixture of secondary and tertiary contacts. This does not belittle the
use of these studies since it is known that the secondary structures
generally form before or meanwhile the tertiary structure
does. Indeed, in general terms, it is believed that there are three
kinds of possible folding mechanisms: i) the hierarchical mechanism
where the secondary structures form before tertiary contacts, ii) the
nucleation-condensation mechanism where a set of secondary contacts
initiates the growth of the native state and iii) the hydrophobic
collapse mechanism where tertiary hydrophobic contacts initiate the
secondary structures.  In all cases, a mix of secondary and tertiary
structures precede the transition state and the precise folding
mechanism may strongly depend on each specific case.

\subsubsection{Designed heteropolymers}

It is numerically possible to design heteropolymers, with non-covalent
random interactions, that show a protein-like behaviour
\cite{shakhnovich,shakhnovich2}. To this end, let us consider a
heteropolymer on a cubic lattice whose sequence is composed by $N$
monomers $m_i$, $i=1...N$. The interaction energy $E_{ij}$ of two
adjacent and non-covalent monomers, $m_i$ and $m_j$, is supposed to be
a random quenched (i.e. fixed during all the procedure) variable with
zero mean value and a variance $1$ --this sets the energy unit.  From
the "residues" $m_i$ and the matrix $E_{ij}$, the following design
procedure leads to an heteropolymer that folds into a compact
structure $S$.  Given a sequence of the monomers defined by their
position along the chain, one permutes two of them (which corresponds
to an exchange mutation in an evolutionary terminology) and accepts
the permutation if the total energy of the compact structure $S$
decreases. At the end of this annealing procedure in the primary
sequence space, one generally gets an heteropolymer that folds quickly
\cite{shakhnovich,shakhnovich2}.  Moreover, at sufficiently high
temperature, two-state behaviour is often observed.  The dynamics is
usually a "coin and crankshaft" Monte-Carlo type with Metropolis
acceptance rate, known to be ergodic \cite{hillorst}.  A hint to
understand the propensity to fold is the presence of an energy gap
between the native state and the lowest (in energy) misfolded states
\cite{shakhnovich}.  However, the existence of a gap is not a
sufficient condition for the molecule to fold since a flat energy
landscape with a single local minimum energy state (i.e. a golf hole
course) does not lead to a fast folder.  As a consequence, it is
reasonable to think that the annealing procedure in the primary
sequence space indirectly designs a funnelled energy landscape and not
only a single thermodynamically stable state.

\begin{figure}
\includegraphics[width=3.7cm,height=4.cm]{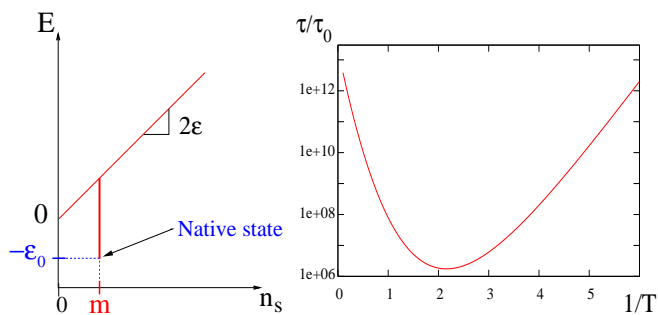}
\includegraphics[width=4.8cm,height=4.cm]{Swanzmod.eps}
\caption{The modified Zwanzig model and the bell-shape curve of the folding time. The original Zwanzig model 
(see Fig. \ref{zwanzig}) can be modified (left picture) in order to take into account a native state shifted
 from the bottom of the valley $n_s=0$. The corresponding folding time as a function of the temperature is reported on 
the right figure. In this figure $\epsilon=0.5$ and $m=5$.
}
\label{bell}
\end{figure}

In numerical studies, one has access at any time to the total number of native contacts, the number of native contacts of
each monomer, the structural overlapping (that quantifies the matching of the relative position
of distant monomers), the end-to-end distance and the gyration radius.
Such an amount of information has led to a good understanding of such systems. 
For instance, it has been shown that the folding rates are correlated with the parameter $\sigma=|T_{\theta}-T_F|/T_{\theta}$ 
\cite{klimov}, 
$T_{\theta}$ and $T_F$ being respectively the Flory coil-to-globule transition and the melting temperatures.
The latter determines the first-order transition between the denatured and the native states. It has also been shown that
the size-dependence of the folding rates is
sensitive to the degree of design \cite{gutin2}. 
Resistance to mutations has
also been studied \cite{broglia}, the main result being that the latter directly depends on
the magnitude of the energy gap. By further adding random interactions to $E_{ij}$, and by
 including hydrophobicity, the phase diagram
in the temperature and denaturant concentration (related to the amount of extra disorder) phase has revealed the presence
of a thermodynamic transition line between compact native structures and coil states but also between native and compact denatured
states as suggested by experiments \cite{finkelstein}. The latter are good candidates to be intermediate states to the folding.

\paragraph{The bell-shape of the folding time}

The above designed heteropolymers lead to a folding time that exhibits
the bell-shape temperature dependence observed in experiments
(Fig. \ref{bell}) \cite{onuchic2}.  The origin of this non-monotonic
behaviour can be twofold. First, it can be due to the roughness of the
energy landscape, which becomes the limiting rate factor when the
thermal energy is on the order of the energy barriers separating the
multiple configurations associated to the denatured state
\cite{bryngelson,onuchic2}.  The simplest corresponding model
describing this scenario is due to Zwanzig \cite{zwanzig}.  By
introducing, in the microscopic time scale of the original model of
Fig.\ref{zwanzig}, a multiplicative Arrhenius factor $\exp(\Delta
E/k_B T)$, one recovers a non-monotonic behaviour for the folding
time. The argument is as follows. At high temperature (entropic regime), the folding
time is large because of the high entropy of the denatured states (see Fig. \ref{zwanzig}). At low
temperature, the folding time is large because of the trapping of
misfolded states whose presence is reflected in the modified
microscopic timescale.  The life-time of the latter is on the order of
$\exp(\Delta E/k_B T)$.  Second, it can also be the manifestation of a
crossover between a regime dominated by entropic effects (high
temperature) and a regime dominated by activated events not related to
any glass transition \cite{gutin}. As an example, let us consider the
picture as proposed by Zwanzig \cite{zwanzig}.  Instead of taking into
account a native state at the bottom of a potential energy valley, we
can define a native state shifted with respect to the bottom of the
valley. If the native state is shifted to a distance $m$
(Fig. \ref{bell}), a calculation that assumes partial equilibration
out of the native state leads to a folding time $\tau=f(m)$
\cite{junier} that has a bell-shape as reported in Fig. \ref{bell}.
In general, the folding time is given by $\tau=g(n_s^*)$ where $n_s^*$
is the average distance between the unfolded state and the native
state. The minimum folding time then corresponds to $n_s^*=m$.  At
high temperatures, the entropy favours large $n_s^*$ whereas at low
temperature, $n_s^* \approx 0$ dominates. In this case, the dynamics
is activated and one finds an Arrhenius law $\tau \sim
\exp(2m\epsilon/T)$.

\begin{figure}
\includegraphics[scale=0.3]{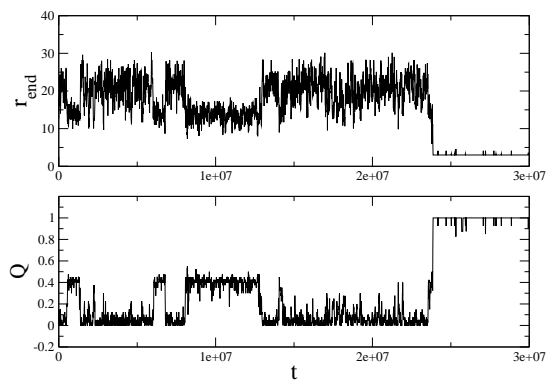}
\caption{Three-state behaviour in a simulation of an heteropolymer on
lattice. In this simulation, the mechanical force $\vec f$ is
incorporated by adding a mechanical energy of the type $-\parallel
\vec f \parallel \times \parallel \vec r_{end-to- end}
\parallel$. This is different from the usual scalar product in order
to prevent the geometrical effects of the lattice (details in
\cite{junier}). The upper panel shows the temporal evolution of the
end-to-end distance and the lower panel shows the corresponding
evolution of the percentage of native contacts.}
\label{three}
\end{figure}

\paragraph{Force-induced transitions}

A few numerical investigations of designed heteropolymer sequences 
in the presence of force have been done. The study in \cite{socci} 
has
revealed a tricky interplay between different
reaction coordinates, e.g. the end-to-end distance and the number of native contacts. This reminds eventual problems
in interpreting the diffusive dynamics in a projected free energy landscape. 

More generally, interesting investigations by Geissler and Shakhnovich \cite{geissler} 
have shown that stretched designed heteropolymers
should behave differently than stretched random heteropolymers. In particular, they argue that
only protein-like sequences would reproducibly unfold and refold at a specific force. Also related to the protein-like
behaviour, it has been shown that a simple stretched polymer at a temperature smaller than its $\theta$-temperature 
(the 
coil-to-globule 
transition) leads at some force to the formation of the $\alpha$-helix secondary structure \cite{marenduzzo}.

\begin{figure}
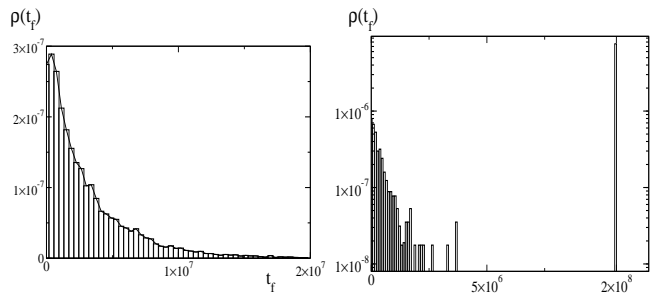

\includegraphics[width=4.2cm,height=3.8cm]{onpathway.eps}
\includegraphics[width=4.2cm,height=3.8cm]{offpathway.eps}
\caption{Left: the distribution of folding times from the intermediate states suggest a case where
only on-pathway states are present. Right: in contrast, off-pathway-states are characterized by a peak at very large time
(here $2\times 10^8$).
This peak actually corresponds to a cut-off in the simulation and would theoretically correspond to an infinite time. 
In this example, one finds $48\%$ of on-pathway states and $52\%$ of off-pathway states.}
\label{onoff}
\end{figure}

Stretched designed heteropolymers on-lattice 
can illustrate the presence of on and off pathway states. 
Indeed, in some conditions of
temperature and force, a three-state behaviour can be observed \cite{junier} --see Fig. \ref{three}. 
We then carried out the force-protocol described in section \ref{inter} to quantify the fraction of misfolded states with 
respect to the on-pathway
states. To this end, each time the system reaches an extension and a percentage of native contacts compatible with
an intermediate state, the force is set to zero and the distribution of folding time from this very moment is computed. 
In small sized systems, fluctuations are important and 
an unfolding event at zero force is always observed. As a consequence, we numerically constrained the system to stay in
the phase space region corresponding to the intermediate state (details in \cite{junier}). In a situation
with only on-pathway states, the distribution is nearly exponential as reported in Fig. \ref{onoff}(left). 
When off-pathway states are present,
the distribution consists of two well-separate contributions (Fig. \ref{onoff}(right)).
The peak observed at very large times is a numerical cut-off time after which we decided to stop the simulation. This
peak corresponds to the off-pathway states.

\subsection{Coarse-grained models}

Coarse-grained simulations allow us to investigate folding kinetics up
to thousands of microseconds. This is not possible by using standard
molecular dynamics due to limited computing power. The underlying
reason to deal with coarse-grained models is the belief that
microscopic details are not determinant to understand the folding
process \cite{baker}.  Usual coarse-grained procedures
\cite{baker,guo,zhou} are inspired from simple G$\bar {\mbox o}$-like
models such as the HP models. These involve an off-lattice dynamics of
only the $C^{\alpha}$ carbons of the polypeptide chain. A typical
model considers three types of carbons (or beads in the literature)
that can be hydrophobic B, hydrophilic L or neutral N
\cite{veitshans}. The energies involved can be divided into two parts:
local and non-local. The local contribution accounts for the covalent
bonds and takes into account harmonic bonds and angle potentials, and
a dihedral angle potential chosen to favour different orientations
according to the surrounding secondary structure
\cite{veitshans,zhou}. Non-local interactions count for the non-bonded
interactions that are responsible for the folding mechanism. They are
usually described by a Lennard-Jones potential and only interactions
between hydrophobic pairs are taken into account. These models allow
to study the statistical properties of the folding process starting
from unfolded configurations, in contrast to the molecular dynamics
simulation. Several results have been obtained about thermal folding
\cite{baker}. For instance, Zhou and Karplus \cite{zhou} have shown
that a wide range of mechanisms could be observed in small helical
proteins just by playing on the energy difference between the native
and the non-native contacts.

In the spirit of the G$\bar {\mbox o}$-like models, let us also mention the use of
mesoscopic elastic models which provide insight on protein dynamics and
folding/unfolding pathways \cite{micheletti}. At variance with other approaches,
the strength of the non-covalent bonds depends on the temperature. This has led to identify 
some interesting differences with random heteropolymers, e.g.
the structural regions involved in slow motions for protein-like models
are much more extended than in random heteropolymers \cite{micheletti}.

In the presence of force, unfolding pathways seem to be mainly related to the structure of the native state \cite{klimov2}.
However, since the study of mechanical properties of proteins is still in its infancy \cite{bensimon, bustamante}, it would be
rather audacious to say that one can in any case deduce the mechanical properties from the structure. Two major 
combined difficulties actually make this investigation difficult: 1) upon mechanical stretching, it is not clear how the network
of forces is distributed inside the protein (for instance, three body interactions are numerous in the native state 
\cite{ejtehadi}), 
and 2) the mechanical properties at the single-residue level are not known.

\paragraph{The instructive RNA case}
A seemingly simpler problem is the one of small RNA hairpins. Indeed, in good solvent conditions,
the native structure corresponds to the secondary structure.
The three dimensional structure is "only" constrained by
the helix arrangement of the different base-pairs. Furthermore, the secondary structure is stabilized by stacking interactions
whose values are well known \cite{jaeger}. 

By adopting a similar coarse-grained model to the one described above,
Hyeon and Thirumalai \cite{hyeon} have studied in detail the
differences between force and thermal induced unfolding. They have
also studied the folding transition by using a force and a thermal
jump protocols. Their study is extremely valuable since such
protocols, especially the force jump experiment, have been realized in
single-protein experiments \cite{fernandez,leeson} and in single RNA
hairpins experiments as well \cite{li}. Their RNA coarse-grained
description is composed of three beads that respectively correspond to
the phosphate, the ribose and the base groups. A dihedral potential
accounts for the right-handed chirality of RNA and a stacking
stabilization potential is incorporated. Hydrophobic interactions
between bases are described by a Lennard-Jones potential endowed with
a distance cut-off and a Debye-Huckel electrostatic potential is
introduced to describe the interaction between the phosphate groups.
They use an overdamped Langevin dynamics and find, as in DNA
experiments \cite{wartell}, that thermal denaturation is due to the
melting of the hairpin where each base-pair fluctuates
independently. In contrast, mechanical denaturation occurs by
sequential unzipping of the hairpin. An interesting prediction is that
the refolding mechanism after a temperature quench should be different
from that after a force quench. In particular, they find that the
folding times upon force quench from stretched states, are much larger
than those upon temperature quench from random states. They explain
this phenomenon by the fact that stretched conditions make the
molecule explore domains of phase space that are inaccessible at high
temperatures by random coil configurations. Interestingly, such a
statement is also valid for proteins and such reported differences are
expected to occur in proteins as well.

\section{Conclusion}

The increasing number of single-protein experiments is providing
insight on the inner details of the protein-folding problem. More
generally, the combination of single-molecule techniques, with and
without force, provide new quantitative results that can be
rationalized with existing theories. In the long term these
experimental results will be useful to better understand the basic
mechanisms underlying many biological processes at the molecular and
cellular level.

The combination of detailed numerical simulations and experiments has
sharpened the theory underlying the propensity of proteins to fold. In
particular, it has confirmed the funnel-like shape of the energy
landscape without excluding well defined steps in the
successive stages of the unfolding/folding transition. The
confrontation of different experimental single-molecule techniques
using mechanical force, on one hand, and fluorescence techniques, on
the other hand, raises new interesting questions about the nature of
the early intermediate states that form during the folding process. The following
questions can now be answered with the recently available techniques:
Under which conditions can
mechanical force stabilize different intermediate states? Is
the intermediate state observed under force the same as the early state
that forms without force? In order to answer these questions, future
design of experiments is needed to obtain structural
information about the intermediate states, the transition states but
also the denatured states.

An alternative approach is to consider protein-like models on-lattice
that show some of these behaviours.  It is then
possible to investigate the different scenarios in a given protein and
clarify whether the two intermediates, that are found with and without
force, are the same or not. In this regard, the mechanical response
studied at various forces, or by applying the force at different
locations along the polypeptide chain, is expected to be different
depending on which scenario is correct. Finally, further simulations
of coarse-grained protein models, in conjunction with experimental
measurements, might lead to improved models that faithfully
reproduce the unfolding/folding pathways of proteins with and without
force.

\begin{theacknowledgments}
We thank M. Palassini for useful comments.
I. J acknowledges financial support form the European network STIPCO,
Grant No. HPRNCT200200319. F. R acknowledges financial support from
the Ministerio de Eduaci\'on y Ciencia (Grant FIS2004-3454 and
NAN2004-09348) and the Catalan government (Distinci\'o de la Generalitat
2001-2005, Grant SGR05-00688).
\end{theacknowledgments}

\bibliographystyle{aipproc}   

\end{document}